\documentclass[aip,jap,reprint]{revtex4-1}
\usepackage{amsmath}
\usepackage{amssymb}
\usepackage{mathrsfs}
\usepackage{graphicx}
\usepackage{bm}
\usepackage{epstopdf}
\usepackage[colorlinks=true,citecolor=blue,linkcolor=blue,urlcolor=blue]{hyperref}
\usepackage{bookmark}
\usepackage{color}
\begin{document}
\title{Optical spectrum of $n$-type and $p$-type monolayer MoS$_2$ in
the presence of proximity-induced interactions}

\author{J. Liu}
\affiliation{School of Physics and Astronomy and Yunnan Key laboratory
of Quantum Information, Yunnan University, Kunming 650091, China}

\author{W. Xu}\email{wenxu$_$issp@aliyun.com}
\affiliation{Micro Optical Instruments Inc., Shenzhen 518118, China}
\affiliation{School of Physics and Astronomy and Yunnan Key laboratory
of Quantum Information, Yunnan University, Kunming 650091, China}
\affiliation{Key Laboratory of Materials Physics, Institute of Solid
State Physics, HFIPS, Chinese Academy of Sciences, HFIPS, Hefei 230031, China}

\author{Y. M. Xiao}\email{yiming.xiao@ynu.edu.cn}
\affiliation{School of Physics and Astronomy and Yunnan Key laboratory
of Quantum Information, Yunnan University, Kunming 650091, China}

\author{L. Ding}
\affiliation{School of Physics and Astronomy and Yunnan Key laboratory
of Quantum Information, Yunnan University, Kunming 650091, China}

\author{H. W. Li}
\affiliation{Micro Optical Instruments Inc., Shenzhen 518118, China}

\author{F. M. Peeters}
\affiliation{Micro Optical Instruments Inc., Shenzhen 518118, China}
\affiliation{Department of Physics, University of Antwerp,
Groenenborgerlaan 171, B-2020 Antwerpen, Belgium}

\date{\today}
\begin{abstract}
In this paper, we examined the effects of proximity-induced interactions
such as Rashba spin-orbit coupling (SOC) and effective Zeeman fields (EZFs)
on the optical spectrum of $n$-type and $p$-type monolayer (ML)-MoS$_2$. The
optical conductivity is evaluated using the standard Kubo formula under Random
phase approximation (RPA) with including the effective electron-electron interaction.
It has been found that there exists two absorption peaks in $n$-type ML-MoS$_{2}$
and two knife shaped absorptions in $p$-type ML-MoS$_{2}$ which are contributed
by the inter-subband spin flip electronic transitions within conduction and valence
bands at valleys $K$ and $K'$ with a lifted valley degeneracy. The optical
absorptions in $n$-type and $p$-type ML-MoS$_{2}$ occur in THz and infrared
radiation regimes and the position, height, and shape of them can be effectively
tuned by Rashba parameter, EZFs parameters, and carrier density. The interesting
theoretical predictions in this study would be helpful for the experimental
observation of the optical absorption in infrared to THz bandwidths contributed by
inter-subband spin flip electronic transitions in a lifted valley degeneracy
monolayer transition metal dichalcogenides (ML-TMDs) system. The obtained results
indicate that ML-MoS$_{2}$ with the platform of proximity interactions make it a
promising infrared and THz material for optics and optoelectronics.
\end{abstract}

\maketitle

\section{Introduction}
In recent years, the discovery of atomically thin two-dimensional (2D)
materials such as graphene and monolayer transition metal dichalcogenides
(ML-TMDs) has been an important and promising field of research in condensed
matter physics \cite{Novoselov04,Mak10,Wang12}. Due to the unique electronic
and optical properties for potential applications in next generation of
high-performance nanoelectronic devices and unique valleytronic features
for information technology \cite{Cao12,Schaibley16,Choi23}, ML-TMDs have
attracted much attention for scientific researches. The electronic
structure of free-standing ML-TMDs is degenerated in $K$ and $K'$ valleys
but with the opposite spin orientations \cite{Xu14}. The valley degeneracy
of ML-TMDs can also be lifted via the exchange interaction induced by the
proximity interaction in the presence of a ferromagnetic substrate \cite{Qi15,Zhang16,Cortes22,Zhao20}.
The proximity-induced 2D ML-TMD based valleytronic system has also led to
the proposal to observe novel optical phenomena such as optical Hall effect
and valley Hall effect \cite{Zhao20,Xiao12}.

The optical and transport properties of ML-MoS$_{2}$ have been theoretically
and experimentally investigated \cite{Li12,Krstajic16,Xiao16,Zhao20,Cao12,Mak12,Radisavljevic13,Mak10,Splendiani10}.
Previous theoretical results have indicated that the splittings of
the conduction and valence bands of ML-MoS$_2$ in the presence of Rashba spin-orbit coupling (SOC) can
introduce optical absorptions in terahertz (THz) to infrared bandwidths \cite{Xiao16}.
The breaking of inversion symmetry at the surface or interface with the
resultant electric field couples to the spin of itinerant electrons is called the
Rashba effect \cite{Soumyanarayanan16}. The Rashba effect can lead to momentum-dependent
splitting of spin bands and would enable spin-flip electronic
transitions \cite{Azpiroz12,Azpiroz13}. It has been shown that the
optical properties such as collective excitations and optical conductivity of
traditional 2D system can be greatly influenced by the Rashba
effect \cite{Xu03apl,Wang05prb,Yuan05,Yang06,Ang14}.
Moreover, the proximity-induced exchange interaction that introduced the
effective Zeeman fields (EZFs) can lift the electronic energy spectrum by
breaking the valley degeneracy \cite{Qi15,Zhao20}. In order to understand the
effect of proximity-induced exchange interaction on the optoelectronic property
of ML-TMDs, it is necessary to examine the roles played by the Rashba SOC and EZFs.

In this study, we evaluate the dependence of longitudinal optical conductivity
on the proximity-induced interactions under the linear polarized
radiation field. The absorption spectrum (optical conductivity) is calculated
via the Kubo formulism under the standard random-phase approximation (RPA) by
including the effective electron-electron interaction. With considering the
contributions of inter-subband spin-flip electronic transitions within conduction
and valence bands in different valleys, the effects of $n$- and $p$-types doping
(for varying carrier density via chemical doping or applying a gate voltage),
EZFs, and the Rashba SOC strength on the optical conductivity of ML-MoS$_2$ at
low temperature are investigated.

The paper is organized as follows. In Sec. \ref{sec:theoretical approach},
the eigenvalues and wavefunctions of ML-MoS$_{2}$ in the presence of
proximity-induced interactions such as EZFs and the Rashba SOC
are obtained by solving the Schr\"{o}dinger equation. The optical conductivity
of ML-MoS$_{2}$ is calculated through the Kubo formulism with the dynamic
dielectric function under RPA. The numerical results of optical conductivity
for different doping types, carrier density, the strengths of EZFs and Rashba SOC
are presented and discussed in Sec. \ref{sec:results}. The concluding remarks are
summarized in Sec. \ref{sec:conclusions}.

\section{Theoretical Framework}
\label{sec:theoretical approach}
In this study, we consider  ML-MoS$_2$ placed on a ferromagnetic substrate
such as EuO or EuS where the proximity-induced interactions can lead to an
enhanced valley splitting and spin-orbit coupling (SOC) \cite{Zhao20}. The
low-energy effective Hamiltonian is written in the form of a $4\times4$ matrix as
\begin{equation}\label{2}
\setlength{\arraycolsep}{0.6pt}
  \begin{split}
&H({\mathbf k})={1\over 2}\times\\
& \left[\begin{array}{cccc}
\Delta+d_\zeta^{c}\ \ & 2A k_\zeta^- \ \ & 0 \ \ & i(1-\zeta )\lambda_R \\
2A k_\zeta^+ \ \ & -(\Delta-d_\zeta^{v}) \ \ & -i(1+\zeta)\lambda_R & \ \ 0 \\
0\ \ & i(1+\zeta)\lambda_R\ \ & \Delta -d_\zeta^{c} & \ \ 2A k_\zeta^- \\
-i(1-\zeta)\lambda_R \ \ & 0 \ \ & 2A k_\zeta^+ \ \ &
-(\Delta+d_\zeta^{v})
\end{array}\right],
  \end{split}
\end{equation}
where ${\bf k}=(k_x,k_y )$ is the electron wavevector along the
2D-plane, $k_\zeta^\pm=\zeta k_x\pm ik_y$, $d_\zeta^{\beta}
=\zeta\lambda_\beta-B_\beta$, and $\beta=(c,v)$.
Here, $\zeta=\pm$ refers to the $K$ ($K'$) valley, $A=at$ with $a$
being the lattice parameter and $t$ the hopping parameter \cite{Cao12}.
The intrinsic SOC parameters $2\lambda_c$ and $2\lambda_v$ are the spin
splitting, respectively, at the bottom of the conduction band and at
the top of the valence band in the absence of the Rashba
SOC \cite{Xiao12,Sun14,Xiao07,Chen99,Zhu11}, $B_c$
and $B_v$ are effective Zeeman fields experienced by electrons in
the conduction band and holes in valence band in the presence of exchange
interaction induced by the substrate. $\Delta$ is the direct bandgap between
the valence and conduction bands\cite{Xiao12,Lu13,Li12}, and $\lambda_R =\alpha_R
\Delta/(2at)$ with $\alpha_R$ being the
Rashba coefficient \cite{Kormnyos14,Slobodeniuk16}. After solving
the Sch\"{o}rdinger equation, one can obtain the eigenvalues
and eigenfunctions of electrons or holes in ML-MoS$_2$.\par

The four eigenvalues $E_{\beta,s}^\zeta(\mathbf{k})$ are the solutions of the diagonalized equation
\begin{align}\label{3}
E^4-A_2E^2+A_1E+A_0=0,
\end{align}
with
\begin{align}
A_2=&{\Delta^2\over 2} +\lambda_R^2
+2A^2k^2+{{d_\zeta^v}^2 +{d_\zeta^c}^2 \over 4},\nonumber\\
A_1=&{\Delta\over 4}({d_\zeta^v}^2-{d_\zeta^c}^2)-{\zeta\lambda_R^2\over 2}
(d_\zeta^v-d_\zeta^c),\nonumber\\
A_0=&\bigl({\Delta^2\over 4}+A^2k^2 \bigr)^2+{\lambda_R^2\over
4}(\Delta+\zeta d_\zeta^c)(\Delta+\zeta d_\zeta^d)\nonumber\\
&-{\Delta^2\over16}({d_\zeta^c}^2 +{d_\zeta^v}^2)-{A^2k^2\over 2}
d_\zeta^cd_\zeta^v+{(d_\zeta^cd_\zeta^v)^2\over 16}\nonumber,
\end{align}
and the corresponding eigenfunctions for electronic states near
the $K$ and $K'$ points are
\begin{equation}\label{4}
|{\bf k};\lambda>={\cal A}_{\beta,s}^\zeta[c_1,c_2,c_3,c_4]e^{i{\bf k}\cdot{\bf r}},
\end{equation}
where ${\bf r}=(x,y)$, $\lambda=(\beta,\zeta,s)$,
\begin{align}
&c_1=i\lambda_R [h_1+4A^2 (1+\zeta)({k_\zeta^-})^2],\nonumber\\
&c_2=-4iA\lambda_Rk_\zeta^- h_2,\nonumber\\
&c_3=2Ak_\zeta^- h_3,\nonumber\\
&c_4=-[(\Delta-2E)^2 -({d_\zeta^c})^2)](\Delta+2E-{d_\zeta^v})\nonumber\\
    &\ \ \ \ \ \ -(1+\zeta)^2\lambda_R^2(\Delta-2E+d_\zeta^c)\nonumber\\
    &\ \ \ \ \ \ -4A^2k^2(\Delta-2E-d_\zeta^c)\nonumber,
\end{align}
and
$${\cal A}_{\beta,s}^\zeta ({\bf k})=(|c_1|^2+|c_2|^2+|c_3|^2+|c_4|^2)^{-1/2}, \nonumber$$
is the normalization factor. Here,
$h_1=(1-\zeta)[\Delta -2E_{\beta,s}^\zeta(\mathbf{k})-d_\zeta^c] [\Delta+2E_{\beta,s}^\zeta(\mathbf{k})-d_\zeta^v]$,
$h_2=\Delta-2E_{\beta,s}^\zeta(\mathbf{k})+\zeta d_\zeta^c$, and
$h_3=[\Delta-2E_{\beta,s}^\zeta(\mathbf{k})+d_\zeta^c][\Delta+2E_{\beta,s}^\zeta(\mathbf{k})-d_\zeta^v]+4A^2 k^2$.

Therefore, we consider the carriers in ML-MoS$_2$ as spin-splitting
2D electron gas (2DEG) in the conduction band or 2D hole gas (2DHG) in the
valence band with a two band mode. Here, we use a simplified form
$E_{s}^\zeta({\bf k})=E_{\beta,s}^\zeta({\bf k})$
and ${\cal A}_{s}^\zeta({\bf k})={\cal A}_{\beta, s}^\zeta({\bf k})$ for both
 conduction and valence subbands. With the energy spectrum of a spin-split
2DEG or 2DHG, the electron density-density ($d$-$d$) correlation function can
be obtained, in the absence of $e$-$e$ screening, as \cite{Chen99,Mishchenko03}
\begin{equation}\label{5}
\Pi_{\alpha}^\zeta (\Omega,{\bf q})
=\sum_{{\bf k}}{A_\alpha^\zeta ({\bf k},{\bf q}){[f(E_{s'}^\zeta({\bf
k}+{\bf q}))-f(E_{s}^\zeta({\bf k}))]}\over E_{s'}^\zeta({\bf k}+{\bf
q})-E_{s}^\zeta({\bf k})+\hbar\Omega+i\delta},
\end{equation}
where
\begin{align}
A_{ss'}^\zeta ({\bf k},{\bf q})=&[{\cal A}_{s'}^\zeta ({\bf k+q})
{\cal A}_{s}^\zeta ({\bf k})]^2\sum_{i=1}^4 c_{is}^{\zeta *} ({\bf k}) c_{is'}^{\zeta}({\bf k+ {\bf
q}})\nonumber\\
&\times\sum_{j=1}^4 c_{js'}^{\zeta*} ({\bf k}+
{\bf q}) c_{js}^{\zeta}({\bf k}),
\end{align}
is the structure factor. Here, $\alpha=(s's)$ is defined for
electronic transition channel from the $s$ branch to the $s'$
branch with $s$=$\pm1$ referring to different spin branches.
$\zeta$=$\pm1$ is for different valley, $\hbar\Omega$
is the excitation photon energy, and ${\bf q}$=$(q_x,q_y)$ is the change of
electron wavevector during an $e$-$e$ scattering event.
$f(x)=[e^{(x-E_F)/k_BT}+1]^{-1}$ is the Fermi-Dirac function
with $E_F$ being the Fermi energy at zero temperature or chemical
potential at a finite temperature. The dynamical RPA dielectric
function hereby writes
\begin{equation}\label{6}
\epsilon(\Omega,q)=1+a_1+a_2+a_3+a_4.
\end{equation}
Here, $(s's)$=$1$=$(++)$, $2$=$(+-)$, $3$=$(-+)$, and $4$=$(--)$
are defined for different transition channels, $(s's)$=$1$ and
$4$ ($2$ and $3$) are for intra-band (inter-subband) transitions,
$a_{s's}$=$-V_{q}\Pi_{s's}^\zeta (\Omega,q)$, and $V_{q}$=$2\pi e^{2}/\kappa q$
with $\kappa$ being the dielectric constant of the material.

The RPA dielectric function can be used to calculate the effective
interaction for optical response with different scattering events.
In the presence of $e$-$e$ interaction, the effective $d$-$d$ correlation
function becomes
\begin{equation}\label{8}
\tilde{\Pi}_{\alpha}^\zeta (\Omega,\mathbf{q})=\Pi_\alpha^\zeta(\Omega,q)/\epsilon(\Omega,q).
\end{equation}
Using the Kubo formula in the absence of electronic scattering centers
(such as impurities and phonons), the optical spectrum or optical conductivity
of a spin-valley-splitting 2DEG (2DHG) can be calculated through \cite{Yuan05,Mishchenko03}
\begin{equation}\label{9}
\sigma(\Omega)=-{\bf }\lim_{ q\rightarrow0}\frac{e^2\Omega}{q^2}
\sum_{\zeta,\alpha}{\rm Im}\ \tilde{\Pi}_{\alpha}^{\zeta}(\Omega,q).
\end{equation}
In the present study, $\sigma(\Omega)$ is induced by current-current correlation via
electron-electron interaction with the external electromagnetic field, which
normally does not change the wavevector for a carrier. In the long-wavelength
limit ($q$$\rightarrow$$0$) and low temperature ($T$$\rightarrow$$0$ K), the
intra-band electronic transitions would not contribute to the optical conductivity.
Moreover, strong optical absorptions can occur via inter-subband spin-flip transitions,
especially for transitions from an energy lower spin branch to an energy higher spin branch
at $K$ and $K'$ valleys.

\section{Results and discussions}
\label{sec:results}

In numerical calculations, we consider the case at low temperature
($T$$\rightarrow$$0$ K) to calculate the optical conductivity and
take the following parameters for ML-MoS$_{2}$ with $A=3.5123$ {\AA}eV,
$\Delta=1.66$ eV, $\lambda_{c}=-1.5$ meV, and $\lambda_{v}=75$ meV \cite{Li12,Ochoa13}.
Usually, the Rashba parameter $\lambda_{R}$ depends on the type of the
substrate and can also be tuned through, e.g., applying a gate voltage\cite{Yao17}.
A large Rashba parameter $\lambda_{R}$ = $72$ meV was found in ML-MoTe$_{2}$ placed
on an EuO substrate \cite{Qi15}. The EZFs factors $B_{c}$ and
$B_{v}$ also depend on the types of substrates \cite{Liang17}.
For $n$-type ML-MoS$_{2}$, we take the spin relaxation time for inter-subband
spin-flip transitions as $\tau_{e}=$ 3  and $\tau_{h}=$ 300 ps for $p$-type
ML-MoS$_{2}$ \cite{Yang13}. With the energy relation approximation, one can
replace the $\delta$ function in Eq. \eqref{5} with a Lorentzian distribution: $\delta(E)\rightarrow(E_{\tau}/\pi)/(E^2+E_{\tau}^{2})$, where $E_{\tau}=\hbar/\tau$
is the width of the distribution \cite{Stille12}. It should be noted that energy
relaxation time is a frequency-dependent parameter and is usually set to a constant
for numerical calculation \cite{Nicol08}. We take the dielectric constants for air,
bare ML-MoS$_{2}$ sheet, and bare EuO substrate as $\kappa_{\mathrm{air}}=1$,
$\kappa_{\mathrm{TMD}}=3.3$\cite{Paul22}, and $\kappa_{\mathrm{EuO}}=23.9$ \cite{Leroux-Hugon72},
respectively. The chemical potential $\mu_{e/h}$ for electrons
in $n$-type and holes in $p$-type ML-MoS$_{2}$ can be determined through
\begin{equation}\label{10}
n_e=\sum_{\zeta=\pm1,s=\pm1}\sum_{\bf k}f(E_s^{\zeta}({\bf k})),
\end{equation}
and
\begin{equation}\label{11}
n_h=\sum_{\zeta=\pm1,s=\pm1}\sum_{\bf k}[1-f(E_s^{\zeta}({\bf k}))],
\end{equation}
respectively. As we know, the longitudinal optical conductivity of ML-MoS$_{2}$
on a ferromagnetic substrate in the presence of proximity-induced interactions can be
measured by the infrared spectroscopy and THz TDS measurement. The theoretical
model in this study provides an excellent platform to examine the optical absorption
property by tuning the parameter such as Rashba parameter $\lambda_R$, EZFs $B_{c}$
and $B_{v}$, and carrier density.

\begin{figure}[t]
\includegraphics[width=8.6cm]{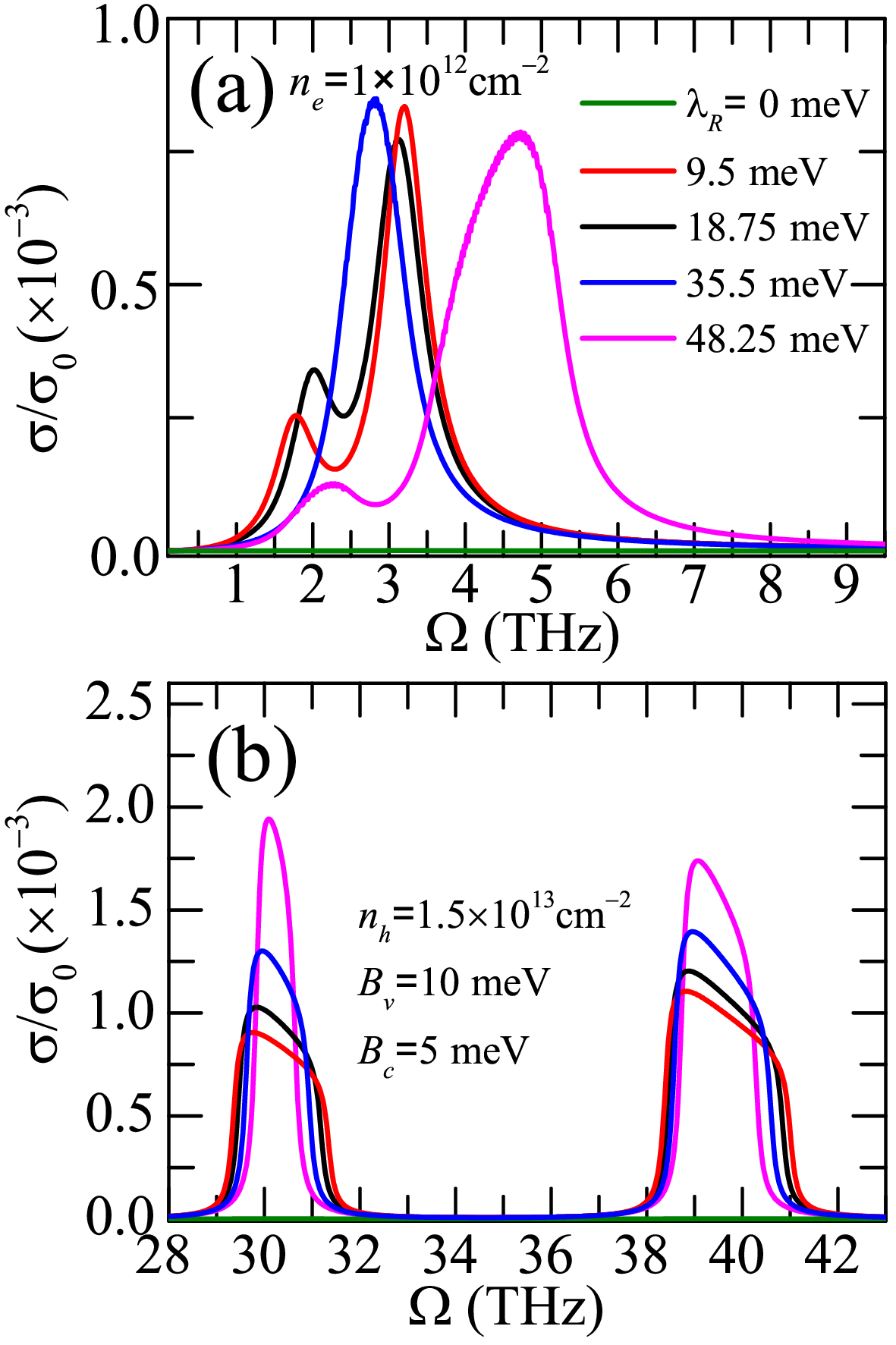}
\caption{Optical conductivity $\sigma(\Omega)$ as a function of
radiation frequency $\Omega$ at a fixed carrier density
(a) $n_{e}=10^{12}$ cm$^{-2}$ for $n$-type and (b) $n_{h}=1.5\times10^{13}$
cm$^{-2}$ for $p$-type ML-MoS$_{2}$ with $B_{v}=10$  and $B_{c}=5$
meV for different Rashba parameters $\lambda_{R}$ as indicated.
Here, $\sigma_0=e^2/(4\hbar)$ is the universal optical conductivity of graphene.}
\label{fig1}
\end{figure}

\begin{figure}[t]
\includegraphics[width=8.6cm]{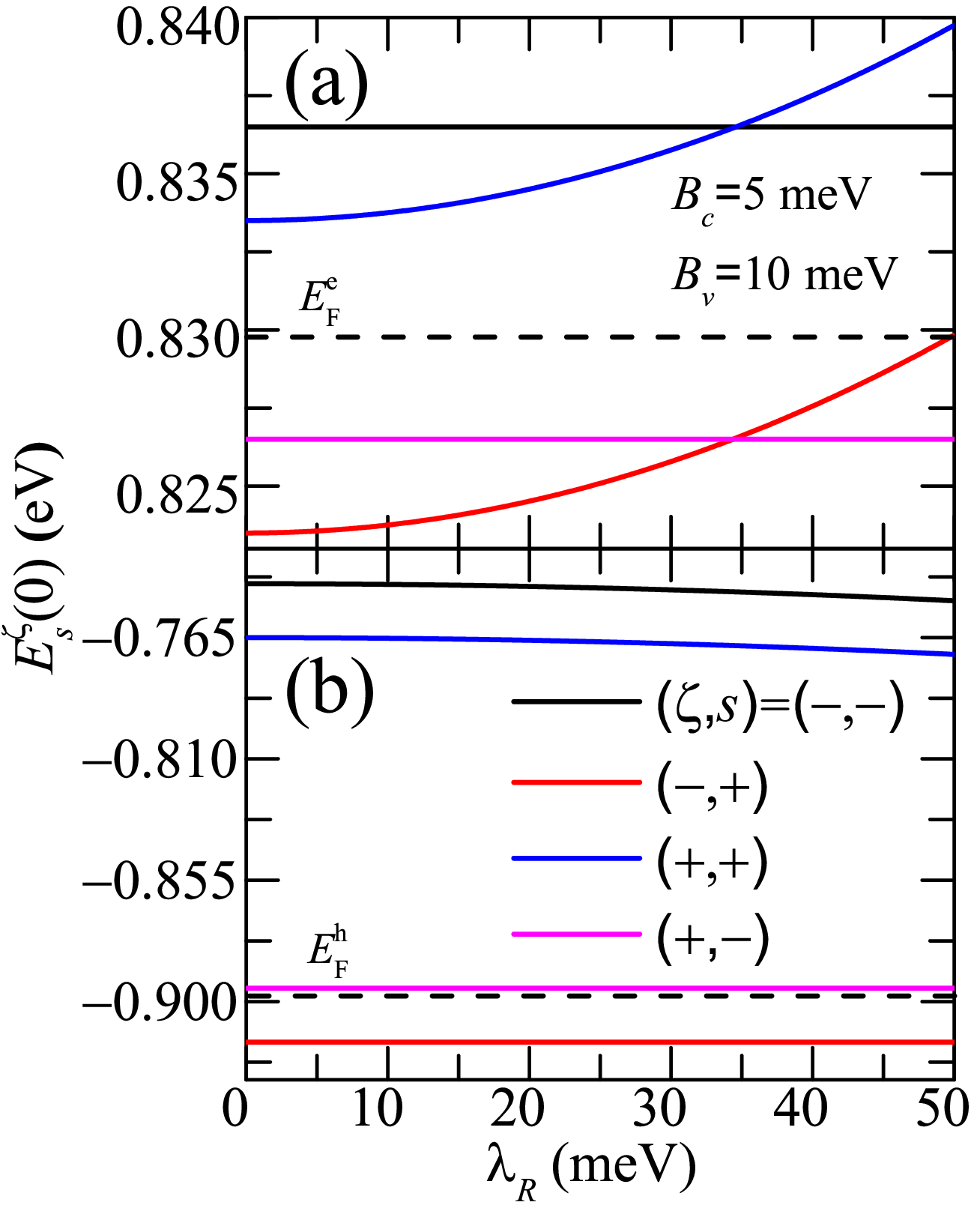}
\caption{(a) The minimum of the four conduction subbands ($\zeta=\pm$,$s=\pm$)
and (b) the maximum of the four valence subbands as a function of Rashba
parameter $\lambda_{R}$ for fixed EZFs $B_{v}=10$  and $B_{c}=5$ meV
at $k=0$.}
\label{fig2}
\end{figure}

\begin{figure}[t]
\includegraphics[width=8.6cm]{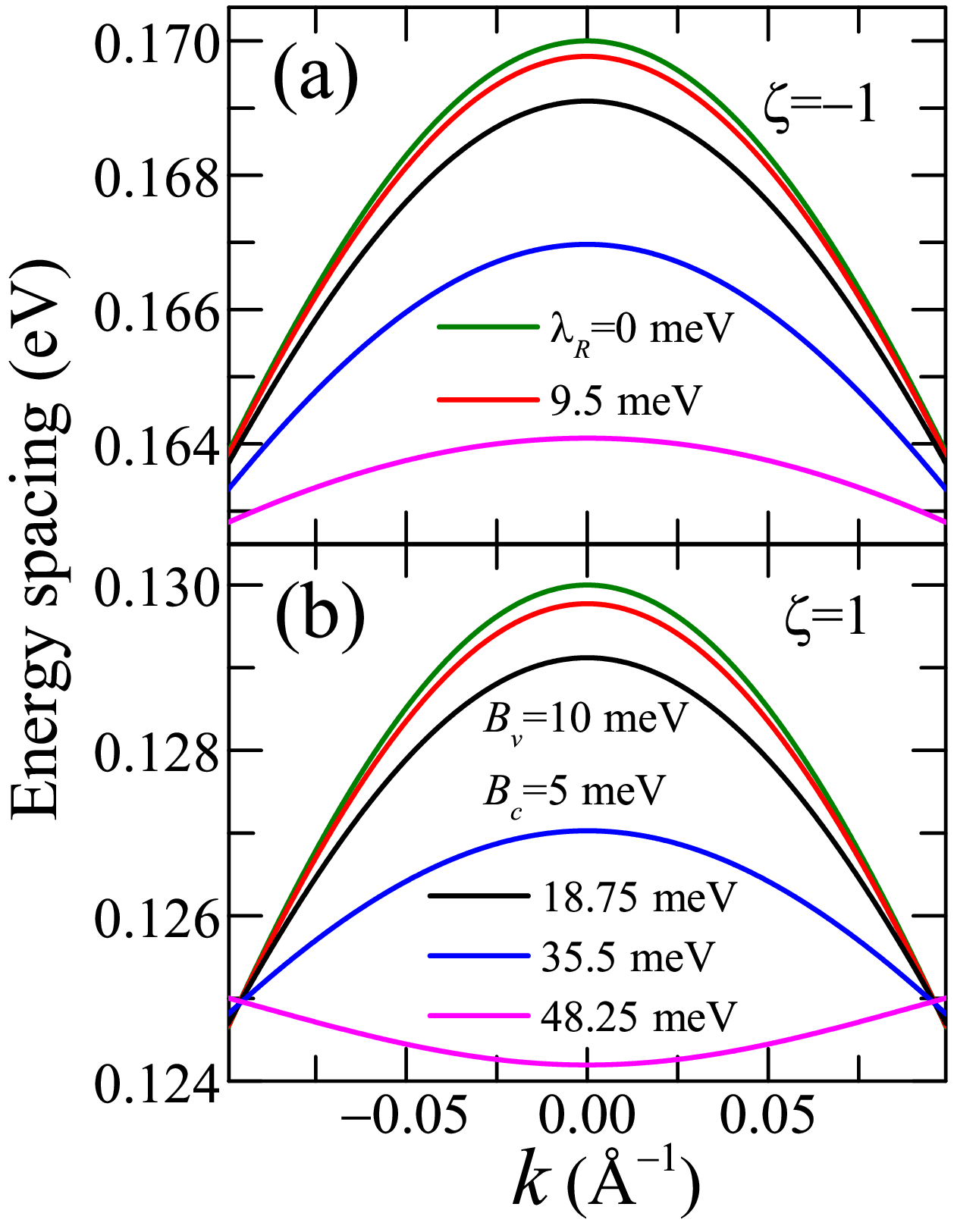}
\caption{The energy spacing between two spin splitting valence subbands
as a function of wavevector $k$ in (a) $K'$ valley and (b) $K$ valley
with different Rashba parameters $\lambda_{R}$.}
\label{fig3}
\end{figure}

In Fig. \ref{fig1}, we plot the optical conductivity of $n$-type and $p$-type
ML-MoS$_{2}$ as a function of radiation frequency $\Omega$ at fixed
carrier density $n_{e}=10^{12}$ cm$^{-2}$ for electrons and $n_{h}=1.5\times10^{13}$ cm$^{-2}$
for holes, $B_{c}=5$ meV, and $B_{v}=10$ meV for different Rashba parameters $\lambda_{R}$.
In Fig. \ref{fig1}(a), we find that there exist two absorption peaks where the
lower left peak is induced by spin-flip electronic transitions within $K$ valley and
the higher right peak is attributed to spin-flip electronic transitions within $K'$
valley. With increasing $\lambda_{R}$, for $\lambda_{R}<35.5$ meV,
the lower left absorption peak blueshifts to higher frequencies and the
higher right peak redshifts to lower frequencies.
For $\lambda_{R}=35.5$ meV, there is only one absorption peak because
the band structure is valley degenerated in this case. While for $\lambda_{R}>35.5$ meV,
the higher peak redshifts to lower frequencies and the lower peak blueshifts to
higher frequencies with increasing $\lambda_{R}$. In Fig. \ref{fig1}(b),
there are two roughly knife shaped spectral absorptions in the infrared regime.
With increasing $\lambda_{R}$, the widths of the knife shaped absorptions
decrease and the heights of the knife shaped absorptions increase. In the absence
of Rashba effect ($\lambda_{R}=0$ meV), we notice that the optical conductivity
approaches to zero because the spin-flip transitions are prohibited in this case.

The interesting findings in Fig. \ref{fig1}(a) can be understood with the help
of Fig. \ref{fig2}(a) where we show the lowest energies at the bottom of four conduction
subbands $(\zeta=\pm,s=\pm)$ as a function of Rashba parameter $\lambda_{R}$ for
fixed EZFs $B_{c}=5$ meV and $B_{v}=10$ meV. With increasing $\lambda_{R}$, the
energy spacing $E_{+}^{+}(0)-E_{-}^{+}(0)$ in valley $K$ becomes larger and the
energy spacing $E_{-}^{-}(0)-E_{+}^{-}(0)$ in valley $K'$ decreases, which would
result in the blueshifts and redshifts of the two absorption peaks.
In Fig. \ref{fig2}(b), we show the highest energies at the top of four valence
subbands $(\zeta=\pm,s=\pm)$ as a function of Rashba parameter $\lambda_{R}$ for
fixed EZFs $B_{c}=5$ meV and $B_{v}=10$ meV. As we can see, the Rashba parameter
affects slightly the top points of them. At a fixed carrier density, we can see
that the Fermi level for electrons/holes depends weakly on Rashba parameter,
which are in line with the situation we had discussed previously \cite{Xiao16}.
Thus, the modification of the band structure by the Rashba effect would change
the energy spacing between spin splitting subbands near the Fermi level and would
result in the tuning of absorption peaks or knife shapes in Fig. \ref{fig1}.
In Fig. \ref{fig3}, we also plot the energy spacing between two spin splitting
subbands in the valence band for $K$ and $K'$ valleys to clearly clarify the Rashba effect on $p$-type sample. The Rashba effect would affect the energy
spacing between the spin splitting valence subbands. With increasing $\lambda_{R}$,
the region of energy spacing near the Fermi level becomes narrower. Thus, the
knife shaped spectral absorptions in Fig. \ref{fig1}(b) also get narrower with
increasing $\lambda_{R}$. The right boundaries of the two knife shape absorptions
correspond to the highest energies in Figs. \ref{fig3}(a)-(b) for the largest
electronic transition energies required by optical absorptions in $p$-type sample.
As we can see, the regions and values of energy spacings in Fig. \ref{fig3}(a)
are larger and higher than those in Fig. \ref{fig3}(b). As a consequence, the
left and right knife shaped absorptions in Fig. \ref{fig1}(b) are contributed
by the spin-flip electronic transitions in valleys $K$ and $K'$, respectively,
and the right knife shaped absorption is more wider than the left one. The
strengths of the knife shaped absorptions also increase with increasing $\lambda_{R}$.
It should be noted that the electronic band structure in ML-MoS$_{2}$ modified
by EZFs and Rashba SOC have some differences from traditional 2DEG or 2DHG system
in the presence of Rashba effect. Thus, the optical spectrum has both commons
and differences with traditional 2DEG and 2DHG \cite{Ang14,Yuan05}. In general,
the Rashba effect can play an important and peculiar role in affecting and tuning
the optical absorptions in THz and infrared bandwidths for $n$- and $p$-type
ML-MoS$_{2}$ in the presence of proximity-induced exchange interaction.

\begin{figure}[t]
\includegraphics[width=8.6cm]{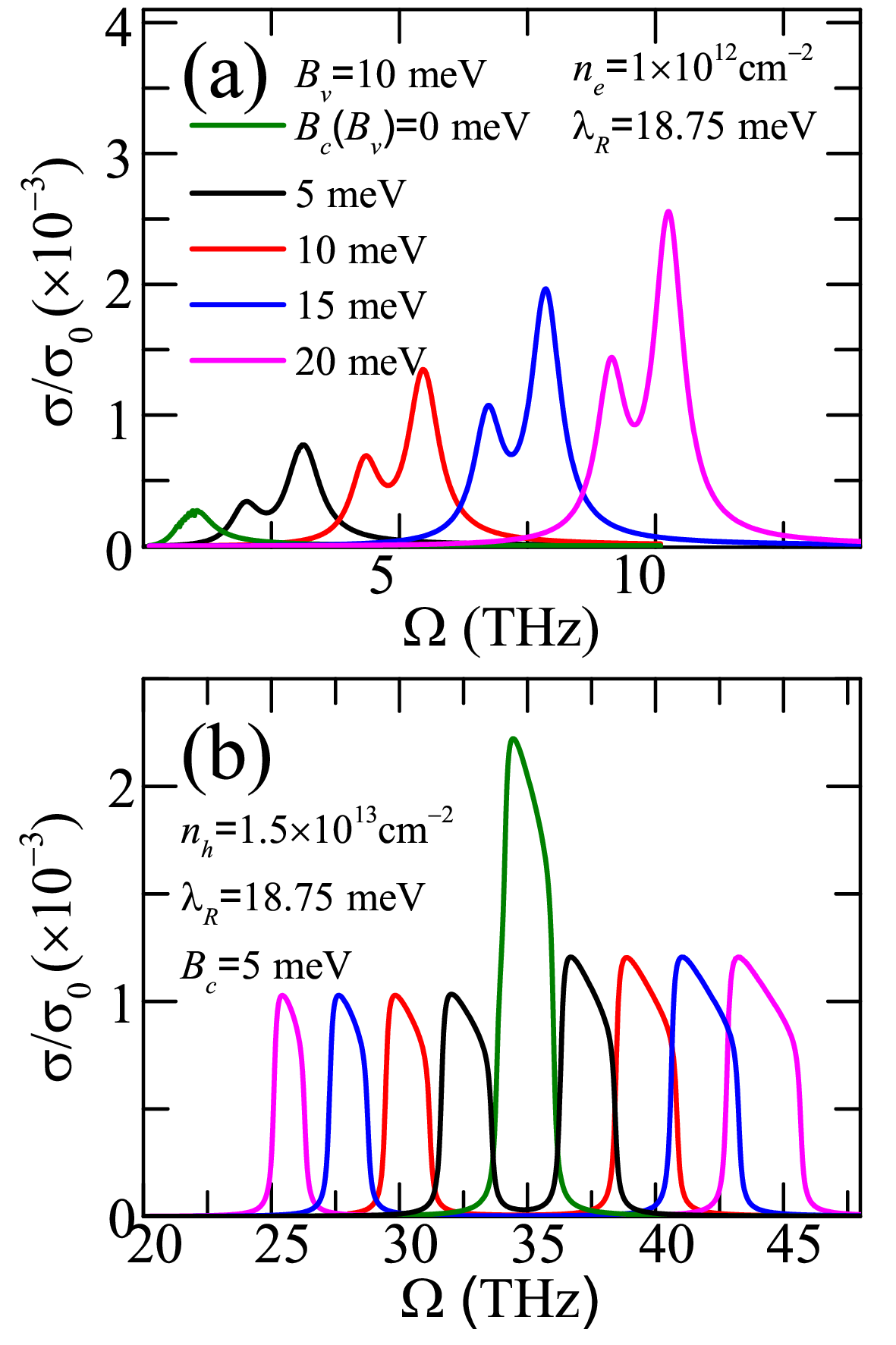}
\caption{(a) Optical conductivity $\sigma(\Omega)$ as a function of radiation
frequency $\Omega$ at a fixed carrier density $n_{e}=10^{12}$ cm$^{-2}$,
EZFs parameter $B_{v}=10$ meV, and Rashba parameter $\lambda_{R}=18.75$ meV for
$n$-type ML-MoS$_{2}$ with different EZF parameters $B_{c}$ as indicated.
(b) Optical conductivity $\sigma(\Omega)$ as a function of excitation
frequency $\Omega$ at a fixed hole density $n_{h}=1.5\times10^{13}$ cm$^{-2}$,
EZFs parameter $B_{c}=5$ meV, and Rashba parameter $\lambda_{R}=18.75$ meV for
$p$-type ML-MoS$_{2}$ with different EZF parameters $B_{v}$ as indicated.}
\label{fig4}
\end{figure}

\begin{figure}[t]
\includegraphics[width=8.6cm]{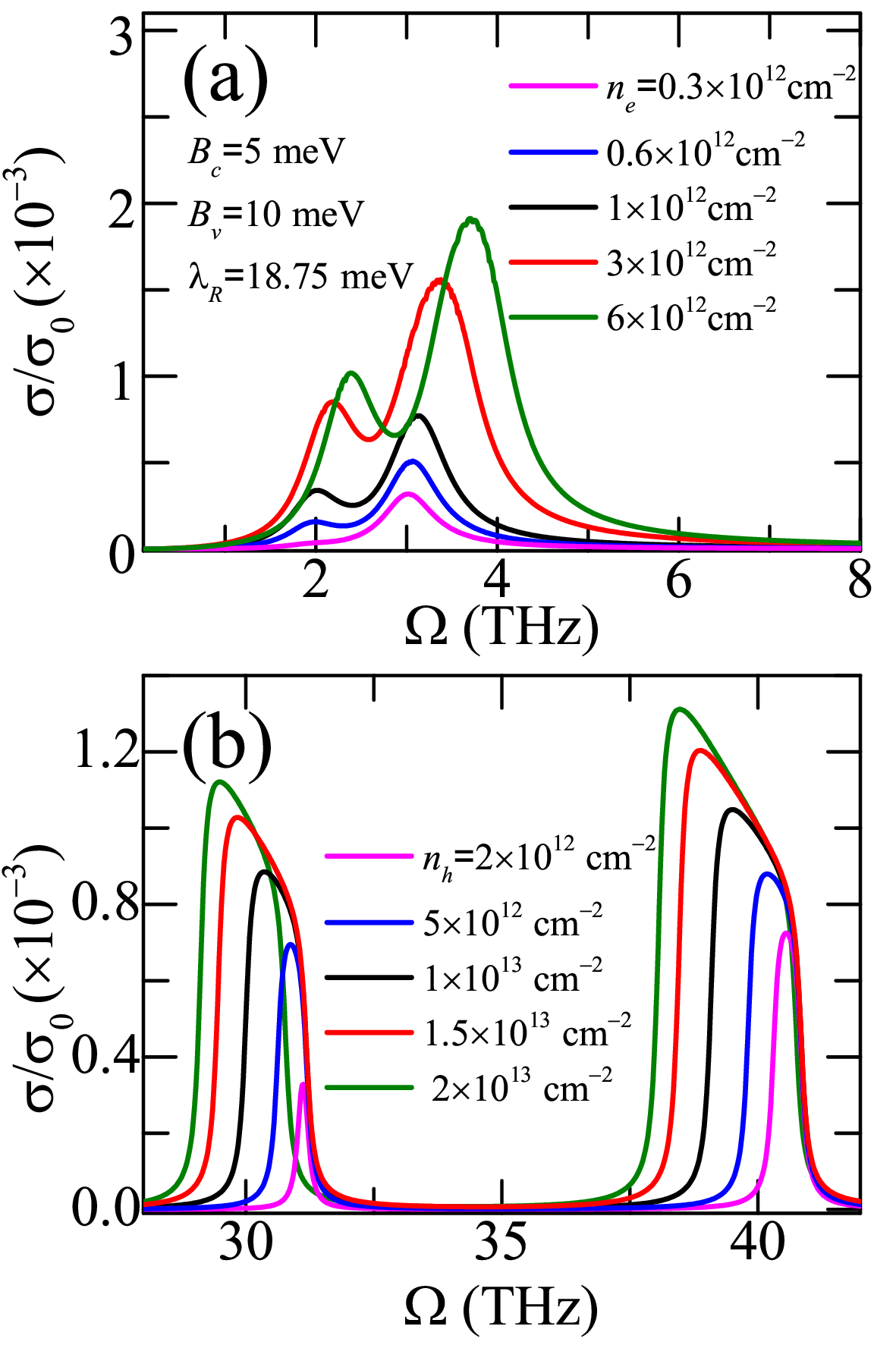}
\caption{Optical conductivity $\sigma(\Omega)$ as a function of radiation
frequency $\Omega$ at fixed Rashba parameters $\lambda_{R}=18.75$ meV, EZF
parameters $B_{c}=5$ meV and $B_{v}=10$ meV for (a) $n$-type ML-MoS$_{2}$ and (b)
$p$-type ML-MoS$_{2}$ with different electron (hole) density $n_{e}$ ($n_{h}$) as
indicated.}
\label{fig5}
\end{figure}

The optical conductivity of $n$- and $p$-type ML-MoS$_{2}$ is shown in
Fig. \ref{fig4} as a function of radiation frequency $\Omega$
at fixed carrier density and $\lambda_{R}$ for different EZFs.
In Fig. \ref{fig4}(a), with a fixed $B_c$= 10 meV, we find that
the width of absorption regime in optical conductivity curve increases
and both of the two absorption peaks blueshift with increasing $B_{c}$.
The strengths of two absorption peaks also become stronger with increasing $B_{c}$.
For $B_{c}=0$ meV, there exists only one peak because the conduction
bands in $K$ and $K'$ valleys are degenerated in this case.
For the optical spectrum for $p$-type ML-MoS$_{2}$ shown in Fig. \ref{fig4}(b),
there is only one knife shape absorption when $B_{v}=0$ meV because of the
valley degeneracy of valence bands in $K$ and $K'$ valleys.
For $B_{v}>0$ meV, there are two knife shape absorptions. With increasing $B_{v}$,
the left knife shape absorption redshifts and the width of it becomes more narrower.
Meanwhile, the right one blueshifts and the width of it becomes more broader.

In Fig. \ref{fig5}, we plot the optical conductivity of $n$-type and $p$-type
ML-MoS$_{2}$ as a function of radiation frequency $\Omega$ for fixed $\lambda_{R}=18.75$ meV,
EFZs $B_{c}=5$ meV and $B_{v}=10$ meV for different carrier densities.
For ML-MoS$_{2}$, $n$-type and $p$-type doping samples can be realized through the field
effect with different source and drain contacts \cite{Chuang14,Radisavljevic13} and
the doping levels can be tuned through, e.g., applying a gate voltage.
Usually, one can reach high carrier density in experiment \cite{Chuang14,Cuong14}
and we choose the carrier density with a magnitude of $10^{12}$ cm$^{-2}$ and hole
density $10^{13}$ cm$^{-2}$ in our numerical calculation. In Fig. \ref{fig5}(a),
we can see that both of the two absorption peaks have blueshifts to higher
frequencies and the strengths of them become stronger with increasing electron
density $n_{e}$. Two knife shaped absorptions in Fig. \ref{fig5}(b) become wider
and  their left boundaries move to the low frequency region with increasing hole
density $n_{h}$. The left boundaries of two knife shape absorptions are redshifts
with increasing carrier density because the chemical potential for holes in $p$-type
sample decreases which allows the spin-flip transitions in the lower frequency regime.
The width of the right knife shaped absorption in Fig. \ref{fig5}(b) is wider than
the left one because the region of energy spacing $E_{-}^{-}(k)-E_{+}^{-}(k)$ in
valley $K'$ is larger than $E_{+}^{+}(k)-E_{-}^{+}(k)$ in valley $K$. The interesting
findings in Fig. \ref{fig5} are due to the Pauli blockade effect with changing the
carrier density \cite{Krenner06}. These theoretical results show that the optical
absorption of ML-MoS$_{2}$ in THz and infrared regimes can also be effectively tuned
by varying the carrier density.

It should be noted that in our present study, the optical spectrum is mainly
determined by electronic transitions through $e$-$e$ interaction at low temperature
which is an ideal case where the electronic scattering mechanisms (e.g., impurities
and phonons) are not taken into account. However, the shape and amplitude of optical
conductivity could be also affected by impurity or electron-phonon scattering with
the modification of self-energy \cite{Li13}.

\section{Conclusions}
\label{sec:conclusions}

In this paper, we have investigated the infrared to THz optical absorption
property of $n$- and $p$-types ML-MoS$_{2}$ in the presence of proximity-induced
interactions such as Rashba SOC and exchange interaction. The
optical conductivity is evaluated using the standard Kubo formula under RPA by
including the effective electron-electron interaction.
We have examined the roles of proximity-induced interactions in affecting
optical absorptions occurring in different valleys for both $n$- and $p$-type ML-MoS$_{2}$.
The main conclusions obtained from this study are summarized as follows.

In the presence of proximity-induced interactions, there exist two absorption peaks
in $n$-type ML-MoS$_{2}$ and two knife shaped absorptions in $p$-type ML-MoS$_{2}$
which are contributed by the inter-subband spin-flip electronic transitions
at valleys $K$ and $K'$ with a lifted valley degeneracy. The height and position
of the absorption peaks in $n$-type ML-MoS$_{2}$ can be effectively tuned by Rashba
parameter, EFZs parameter $B_{c}$, and electron density. The width, height, and
position of the knife shaped absorptions in $p$-type ML-MoS$_{2}$ depend strongly
on the Rashba parameter, EFZs parameter $B_{v}$, and carrier density. These features
in optical conductivity curves can be explained by the electronic transition channels
and the modification of conduction and valence bands by the proximity induced
Rashba SOC and EZFs. The obtained results suggest that ML-MoS$_{2}$ in the presence
of Rashba SOC and EZFs has a wide tunable optical response in the infrared to
THz radiation regimes. The optoelectronic properties of $n$-type and $p$-type
ML-MoS$_{2}$ in the presence of proximity-induced interactions can be effectively
tuned by the carrier density, Rashba parameter, and EZFs which makes ML-MoS$_{2}$
a promising infrared and THz material for optics and optoelectronics.
The obtained theoretical findings can be helpful for the understanding of
optoelectronic properties of ML-MoS$_{2}$. We hope the theoretical
predictions in this paper can be verified experimentally.

\section*{ACKNOWLEDGMENTS}
This work was supported by the National Natural Science foundation
of China (NSFC) (Grants  No. U2230122, No. U2067207, No. 12364009,
and No. 12004331), Shenzhen Science and Technology Program (Grant
No. KQTD20190929173954826), and by Yunnan Fundamental Research
Projects (Grants No. 202301AT070120, and No. 202101AT070166). Y.M.X.
was supported through the Xingdian Talent Plans for Young Talents
of Yunnan Province (Grant No. XDYC-QNRC-2022-0492).

\end{document}